# High-pressure hydrothermal growth and characterization of $Sr_3Os_4O_{14}$ single crystals

N. D. Zhigadlo

*CrystMat Company, CH-8037 Zurich, Switzerland*

**Abstract**

Single crystals of the novel strontium osmate $Sr_3Os_4O_{14}$ have been grown by the hydrothermal method using opposed anvil high-pressure and high-temperature technique. The reaction took place in sealed gold capsules at 3 GPa and a temperature of 1100 °C, with water acting as a solvent. The employed method yields up to 1 mm crystals with quite uncommon double-terminated morphologies. The crystal structure was identified as tetragonal by single-crystal X-ray diffraction, with lattice parameters $a$ = 12.2909(8) Å and $c$ = 7.2478(5) Å. The structural analysis suggests *P*4$_2$*nm* or *P*4$_2$/*mnm* as a possible space group. In general, the structure belongs to the pyrochlore type and is composed of a network of symmetrically arranged $OsO_6$ octahedra. Resistivity measurements evidence a metallic behavior, accompanied by a temperature-independent paramagnetism. Heat capacity measurements reveal a slightly enhanced value of the Sommerfeld coefficient $\gamma$ = 34 mJ/mol K$^2$. Superconductivity has not been observed down to 2 K.

E-mail address: nzhigadlo@gmail.com (N.D. Zhigadlo); https://crystmat.com

1. Introduction

Pyrochlores belong to a numerous family of crystalline materials with a wide range of technologically important functional properties and a large structural diversity [1,2]. The common pyrochlore structure is of the type $A_2B_2O_6O'$, often written as $A_2B_2O_7$, where $A$ is a trivalent cation and $B$ a tetravalent cation ($A^{3+},B^{4+}$). These structures, as well as the ($A^{2+},B^{5+}$) combination, have been extensively studied over the past decades due to their promising properties for a wide range of applications. They frequently have a three-dimensional skeleton built of $BO_6$ octahedra and crystallize in a cubic pyrochlore structure with space group $Fd$-$3m$. In many of these structures the materials adopt very different compositions with a wide range of constituent ratios, giving rise to specific phases under proper synthetic conditions. This feature leads to an almost unlimited numbers of possible compositions within the metal-oxide family of materials.

Pyrochlore oxides and compounds containing a $5d$ element, such as Os, have drawn considerable attention in recent years due to the spatial extension of $5d$ orbitals, the modest electron correlations, the large spin-orbit coupling, and the propensity to form flat bands, a manifestation of frustrated hopping [3]. These effects give rise to exotic properties, such as metal-insulator transition, superconductivity, topologically-protected states of mater, and frustrated magnetism. The unexpected discovery of superconductivity in the $Cd_2Re_2O_7$ [4] and $AOs_2O_6$ series ($A$ = Cs, Rb and K) [5-9] pyrochlores, a Slater-like transition in the perovskite oxide $NaOsO_3$ [10], a ferroelectric-like transition in the metallic oxide $LiOsO_3$ [11], and an unusual Mott insulating state in $Sr_2IrO_4$ [12], represent unique features of the $5d$ oxides. The double-perovskite osmium oxides, when synthesized as thin films, also exhibit a rich electromagnetic behavior, including a remarkable ferromagnetic transition above 1000 K reported in $Sr_3OsO_6$ [13]. At the same time, $Sr_3OsO_6$ behaves differently when synthesized under high pressure conditions [14].

Not many $5d$ oxides are known due to the challenging of stabilizing complicated structures with a fixed stoichiometry and the metastability of their phases. Thus, the field of $5d$ oxides, in particular of osmium, remains substantially unexplored and holds great promise for new technological advancements. So far, most osmium oxides have been synthesized in polycrystalline form and only a few of single-crystal growths have been reported [15]. For instance, single crystals of $Cd_2Os_2O_7$ required heating a week, while risking human exposure to the hazardous $OsO_4$. In recent years, high pressure was found to be very effective in growing single crystals of NaOsO$_3$, LiOsO$_3$, and Na$_2$OsO$_4$ [15], by reducing the heating time to a few hours. Importantly, the high-pressure approach is helpful in lowering the risk of OsO$_4$ exposure, since crystal growth attempts are carried out in a closed atmosphere. Thus, this method offers an efficient way to grow single crystals of osmium oxides and to search other pyrochlore compounds in related systems.

Our attempts to synthesize strontium osmate Sr$_2$Os$_2$O$_7$ under high pressure produced instead the unique compound Sr$_3$Os$_4$O$_{14}$, which is another member of the osmate oxides related to the Sr-Os-O system. It should be noted, that the following strontium oxido-osmates have so far been reported in the ternary Sr-Os-O system: SrOsO$_3$ [16], SrOsO$_4$ [17], Sr$_2$OsO$_5$ [18], Sr$_2$Os$_3$O$_5$ [18], Sr$_2$Os$_2$O$_{6.4}$ [19], Sr$_3$OsO$_6$ [14], Sr$_5$Os$_3$O$_{13}$ [20], Sr$_7$Os$_4$O$_{19}$ [18], Sr$_8$Os$_{6.3}$O$_{24}$ [21], Sr$_9$Os$_5$O$_{23}$ [22], and Sr$_{11}$Os$_4$O$_{24}$ [23]. The results of the present study provide further insight into the high-pressure hydrothermal growth, which can be useful to and serve as a guide for the hydrothermal way of producing other intriguing materials.

## 2. Experimental details

Single crystals of Sr$_3$Os$_4$O$_{14}$ were grown using a high-pressure, high-temperature opposed-anvil method. The pressure generation mechanism is based on the compression and confinement of lithographic stone between special working surfaces of a pair of opposed anvil-type dies. The high-pressure module is devised so as to transform the straightforward motion of a hydraulic jack into the inward motion of the anvils. During compression the lithographic stone squeezes out into the spaces between the anvils until the friction between the pressure medium and the anvils balances the pressure generated inside the sample assembly. Using Bi and PbSe as calibrants, the load-pressure relation derived from independent tests was used to estimate the pressure applied to the sample. The temperature of the sample was determined by the pre-calibrated relationship between the applied electrical power and the measured temperature in the cell. The high-pressure method was successfully used earlier by us to grow crystals of various superconducting [24-27] and magnetic materials [28,29], diamonds [30], cuprate oxides [31], pyrochlores [32], *Ln*Fe*Pn*O (*Ln*: lanthanide, *Pn*: pnictogen) oxypnictides [33], as well as a wide range of other compounds [34,35].

The cross-sectional view of the high-pressure sample cell assembly used in the experiments is schematically illustrated in Fig. 1. The furnace is composed of a graphite tube in contact with two anvils through Mo, Ta, Fe and Cu conductive metals. The heater is separated from the lithographic stone by disks made of a 50% NaCl and 50% $ZrO_2$ powder mixture. The initial mixture was sealed inside an Au cylindrical capsule 8.0-mm long and with a 6.8-mm diameter. The capsule was then placed inside a boron nitride (BN) crucible before being mounted into a lithographic stone. The entire crystal growth process lasted no longer than five hours. More information regarding the apparatus design and experimental setup can be found elsewhere [30]. *Caution*: When working with osmium-containing products outside the glovebox, appropriate personal protection equipment should be used because osmium can adopt multiple oxidation states, of which $OsO_4$ is both volatile and poisonous. Closed crucibles of adequate thickness should be used to withstand the vapour pressure of $OsO_4$ during high temperature synthesis. When the synthesis is complete, Au crucibles must be opened in a well-ventilated fume hood to prevent any potential contact with the volatile $OsO_4$.

The general morphology and dimensions of the grown crystals were assessed using an optical microscope (Leica M 205C). Energy-dispersive X-ray spectroscopy (EDX) was used to determine the chemical composition. Structural properties of the $Sr_3Os_4O_{14}$ single crystals were investigated at room temperature with a Bruker SMART X-ray single crystal diffractometer. Four-point resistivity measurements were performed in a Quantum Design Physical Property Measurement System (PPMS). Magnetic measurements between 2 and 300 K in an applied magnetic field of 1 T were performed using a Magnetic Property Measurement System (MPMS) by Quantum Design. For the specific-heat studies, a collection of single crystals was measured in a PPMS apparatus using the adiabatic relaxation technique.

3. **Results and discussion**

Our original idea was to use the hydrothermal method to synthesize the strontium osmate $Sr_2Os_2O_7$ pyrochlore compound as a potential superconductor. A stoichiometric mixture of $2SrO + 2OsO_2 + AgO$, consisting of the initial powdered ingredients SrO (Aldrich, 99%), $OsO_2$ (Alfa Aesar, 99.99%), and AgO (Aldrich, 99%) with addition of 5wt% of distilled $H_2O$, was therefore thoroughly mixed in an agate mortar inside a glove box before being placed inside an Au cylindrical capsule. The capsule was mechanically sealed inside the glove box using a manual press before being placed into the BN crucible. Finally, the entire high-pressure reaction cell was assembled and inserted inside a lithographic stone gasket (soft $CaCO_3$-based material), which served as the pressure transferring medium (Fig. 1). When a sample is squeezed in a high-pressure device, the applied pressure causes the final, complete closure of the capsule. The hydrothermal crystal growth under high pressure and high temperature offers several advantages over the conventional method of solid-state synthesis. Supercritical water acts as a solvent and enables the effective transport of reactants to grow crystal surfaces, thus allowing the rapid development of crystals. Elevated temperatures and pressures can improve the solubility of reactants and facilitate the formation of dense phases. A slight temperature gradient within the reaction capsule can also partially facilitate crystal growth under high pressure and high temperature in the presence of an appropriate amount of flux.

The typical crystals growth process was carried out in the manner outlined below. A pressure of 3 GPa was applied at room temperature. While the pressure was kept constant, the samples were heated during 1 hour to 1100 °C and then maintained there for 2 hours. After that the samples were cooled to room temperature in 2 hours and the pressure was released. The weight of the Au capsule was the same before and after the experiment, thus implying no material loss. Since both ends of the capsule were significantly expanded, this suggests the development of a considerable pressure inside the capsule. Cutting up the Au capsule revealed a large amount of water in addition to the high-pressure products, thus proving that the crystals were grown hydrothermally. The crystals were sonicated in ethanol to separate them from the remaining mass and then air-dried. As a main product, large quantities of thick crystals with a green colour and tapering edges up to 1 mm in length were obtained. An EDX investigation revealed that the Sr/Os ratio was 0.75, leading to the composition $Sr_3Os_4O_{14}$. The latter was subsequently verified by X-ray refinement. Besides the Sr, Os, and O elements, no other chemical elements were found. The actual chemical composition is determined to be stoichiometric in terms of the metal atoms. Furthermore, we can reasonably assume that the compound is also stoichiometric in oxygen because perovskite-like Os oxides are typically oxygen stoichiometric according to several neutron diffraction studies [36]. The preservation of oxygen stoichiometry might be due to the chemical nature of Os. The specified synthesis conditions provided a high yield of $Sr_3Os_4O_{14}$ crystals, while alternative starting materials, reaction temperature, and time led to the appearance of other phases. A lower synthesis temperature implies smaller crystals, which stick closely together.

A collection of $Sr_3Os_4O_{14}$ crystals is shown in Fig. 2. The macroscopic morphology of hydrothermally produced crystals displayed a remarkable diversity. Interestingly, most of the crystals have two termination ends (marked 1 in Fig. 2), and some present penetration twinning (marked 2 in Fig. 2). Double terminated crystals are quite rare specimens and can occasionally be found within minerals. Such crystals frequently form within free-floating pockets of slowly evaporating liquid during their natural growth process, producing perfectly formed crystals with termination on both sides. This can occur under specific geological circumstances or when crystals form in solutions without any physical barriers. Conversely, a single-terminated end indicates that crystal growth either started on a surface that prevented it to form another terminal end, or that it was interrupted after one end formed due to lack of growth space. For example, most minerals are typically terminated only from one side, since they form out of a hot liquid solution, be it magma or mineral-rich water, when it begins to cool. As the solution cools, the dissolved elements bond together to become more stable, a process that reiterates itself until the crystal forms. Our synthetic hydrothermal high-pressure and high-temperature conditions in some way resemble the growth processes that occurs in nature. However, in nature this reaction normally takes even tens of years at relatively low temperatures (50-70 °C), in a classical hydrothermal process several days, while in our case it takes only a few hours. To our great surprise, in our specific conditions, where the sample space is extremely constrained, the grown crystals invariably exhibit double terminations. This might be caused by a solute-rich water solution, as confirmed by the fact that both ends of the reactive Au capsule were significantly expanded, a condition which most likely allows the development of double terminated crystals. In summary, the high-pressure, high-temperature hydrothermal approach can quickly recreate in the lab the processes that take many years to emerge in specific geological environments.

The crystal structure of $Sr_3Os_4O_{14}$ determined using single crystal X-ray diffraction was identified as tetragonal, isostructural to $Pb_3Nb_4O_{12}F_2$ [37], $Ba_2SrMg_4F_{14}$ [38], $Ba_{2.2}Ca_{0.8}Mg_4F_{14}$ [39], with lattice parameters $a$ = 12.2909(8) Å and $c$ = 7.2478(5) Å. The structural view along the [001] direction clearly displays a distinctive tunnel motif shared by numerous pyrochlore oxides. The data analysis suggests $P4_2nm$ or $P4_2/mnm$ as potential space groups. Several models were tested, however no single model could resolve all the discrepancies, since all refinements exhibit a high degree of correlation among the numerous parameters. Similar issues were previously encountered during the refinement of the alike compounds $Sr_4Ru_{3.05}O_{12}$ [40] and $Sr_8Os_{6.3}O_{24}$ [21]. Consequently, we consider the structural model presented here as averaged, yet still close to the actual structure. The above difficulties suggests that, under high-pressure conditions, $Sr_3Os_4O_{14}$ may crystallize into multiple structures. Alternatively, crystal twining may impair the refinement. Based on our structural analysis, it is clear that a conclusive investigation should be carried out utilizing single crystals grown by alternative techniques.

The crystal structure of $Sr_3Os_4O_{14}$ along the [001] direction is shown in Fig. 3. In general, the structure can be classified as a pyrochlore type. Here, twisting chains of $OsO_6$ octahedra, linked to each other, are directed along the face diagonals of the cubic unit cell, forming a symmetrical framework. In this structure, the chains develop only in one direction. In *a* and *b* directions the chains are broken, *i.e.*, only fragments of two octahedra connect chains into a three-dimensional framework. This arrangement of structural fragments creates another type of channels, running in the same direction, namely along the *c*-axis. The presence of terminal oxygen atoms at the Os(2)O$_6$ octahedra and the size of channels suggests the possibility that small cations may be intercalated in them. The novel osmate Sr$_3$Os$_4$O$_{14}$ exhibits mixed-valency with an average osmium oxidation number of 5.5+, in accordance with the conventional oxidation states of Sr$^{2+}$ and O$^{2-}$ and the requirement of charge neutrality.

Temperature-dependent resistivity measurements, $\rho(T)$, indicate a metallic character with a low temperature resistivity of ~ 340 μΩ cm (Fig. 4). The temperature dependence of the Sr$_3$Os$_4$O$_{14}$ magnetic susceptibility $\chi$ shows an almost temperature independent behavior above 100 K. At lower temperature an upturn is seen and the sample becomes paramagnetic (Fig. 5). No magnetic transitions were observed during these measurements. A broad and modest hump was visible at about 47 K, similarly to one seen in Sr$_7$Os$_4$O$_{19}$ [18], although no corresponding anomalies were found in the measurements of resistivity and specific heat. Overall, the magnetic properties were quite similar to those seen in certain other 5*d* oxides, regardless of the number of 5*d* electrons. The lack of a long-range magnetic order in Sr$_3$Os$_4$O$_{14}$ seems quite common among the tunnel-like structures of 5*d* oxides. Note also, that no superconductivity was found down to 2 K.

Fig. 6 shows the temperature dependence of specific heat divided by temperature, $C_p/T$, for Sr$_3$Os$_4$O$_{14}$ single crystals. Over the 2 to 300 K temperature range, $C_p$ varied monotonically with no evident irregularity. The low-temperature region is plotted in the $C_p/T$ vs. $T^2$ form in the inset of Fig. 6 and analyzed by using the Debye model. The relevant fits produced a Sommerfeld coefficient, $\gamma = 34.3$ mJ/molOs K$^2$.

### 4. Conclusions

This study presented the application of an opposed anvil-type high-pressure, high-temperature technique for the hydrothermal growth in a gold container of novel strontium osmate Sr$_3$Os$_4$O$_{14}$, derived from the Sr-Os-O system. The employed method produces crystals with dimensions of about 1 mm and peculiar morphologies in the form of double terminated end points. The crystals crystallize in a pyrochlore-like structure, where the OsO$_6$ octahedra form a symmetrical network. The compound is metallic - as evidenced from resistivity measurements - and superconductivity was not observed down to 2 K. The results and the methodology presented in this work can potentially be used in physical mineralogy to better understand the processes by which crystals form naturally in an aqueous medium at high pressures and temperatures.

**CRediT authorship contribution statement**

N. D. Zhigadlo: Conceptualization, Methodology, Investigation, Data curation, Formal analysis, Visualization, Writing-original draft, Writing - Review & Editing.

**Data availability**

Data will be made available on request.

**Declare of Competing Interest**

The author declare that he has no known competing financial interests or personal relationships that could have appeared to influence the work reported in this paper.

**Acknowledgements**

The author would like to thank his former coworkers Z. Bukowski, M. Brühwiller, and M. Tortello from the Laboratory for Solid State Physics, ETH Zurich for their support, helpful assistance, and valuable discussions. He is also grateful to T. Shiroka for reading the manuscript critically and for offering insightful comments.
This work is dedicated to the memory of our wonderful colleague Zbigniew Bukowski, who sadly passed away recently.

**Figures**

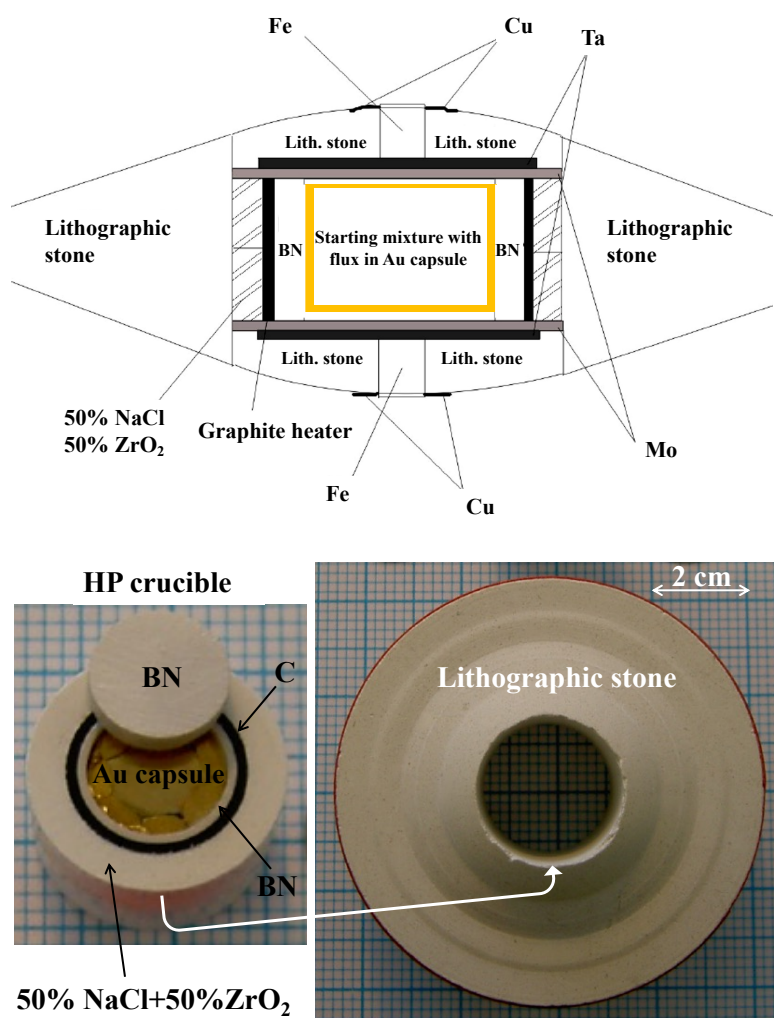

**Fig. 1.** Upper panel. Schematic representation of the cross-sectional view of the high-pressure cell assembly. The furnace is made of a graphite tube that connects to two anvils through conductive materials made of Mo, Ta, Fe and Cu. The initial mixture and solvent are contained in the Au capsule, which is inserted into the BN crucible. Lower panel. A photo of the high-pressure reaction cell before its insertion into the lithographic stone disc.

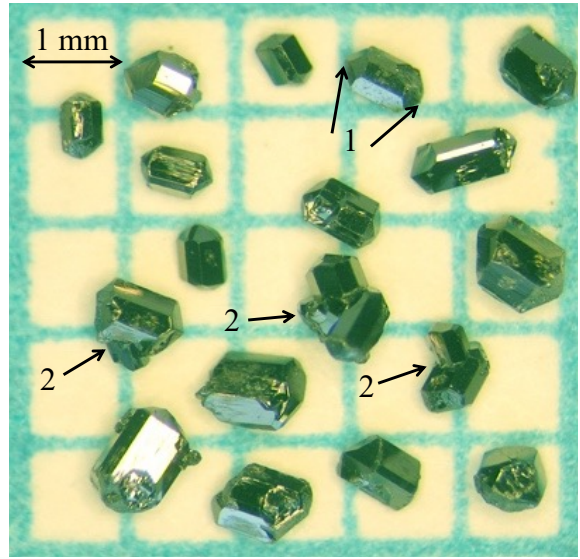

**Fig. 2.** Optical image of $Sr_3Os_4O_{14}$ crystals grown hydrothermally under high pressure and high temperature. The crystals have well-shaped structural morphology with various appearances. Most crystals are double terminated, i.e., they have two ends (marked as 1) and some of them exhibit also penetration twinning (marked as 2).

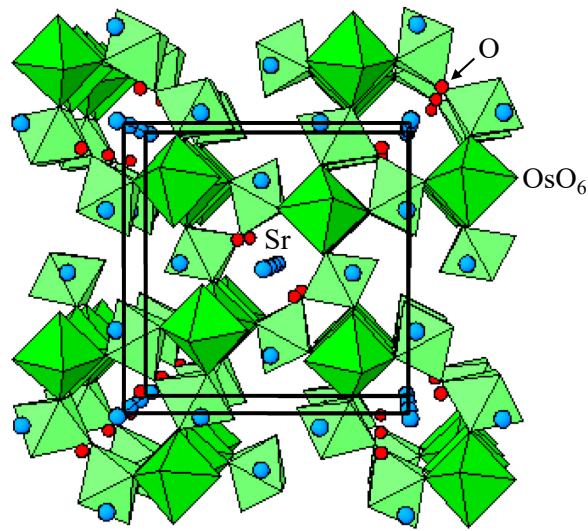

**Fig. 3.** View of the $Sr_3Os_4O_{14}$ crystal structure along the [001] direction. Colour codes: green, octahedra – $OsO_6$; blue, spheres – Sr; red, spheres – O. Solid lines indicate the tetragonal unit cell.

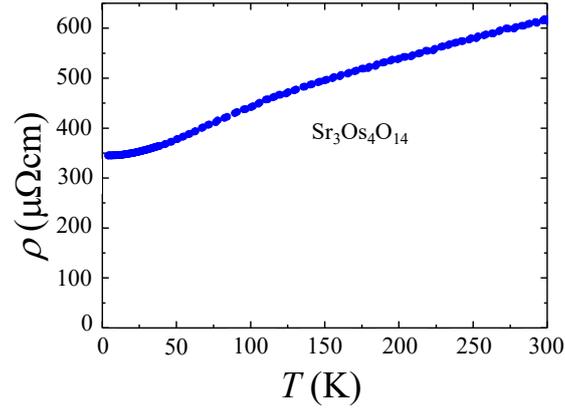

**Fig. 4.** Temperature dependence of the electrical resistivity of a single crystal of $Sr_3Os_4O_{14}$.

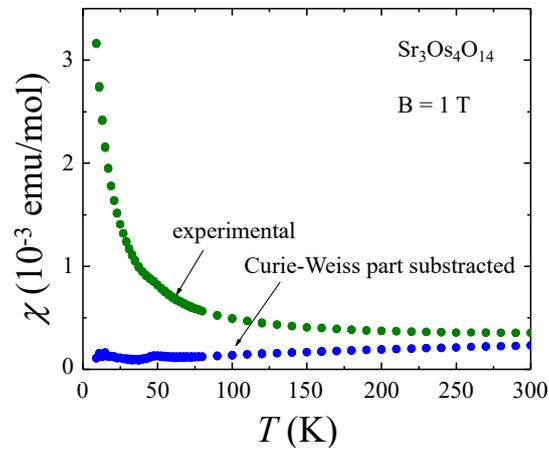

**Fig. 5.** Temperature dependence of the magnetic susceptibility $\chi$ of $Sr_3Os_4O_{14}$ measured in a magnetic field of 1 T.

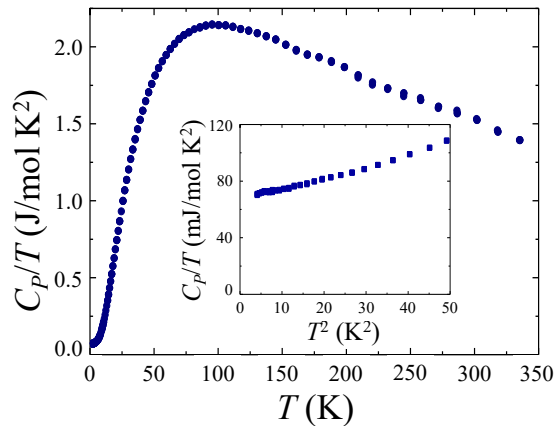

**Fig. 6.** Temperature dependence of the specific heat divided by temperature, $C_p/T$, for $Sr_3Os_4O_{14}$ single crystals. The same $C_p/T$ data are shown in the inset as a function of $T^2$.